\documentclass[3p]{elsarticle}
\UseRawInputEncoding
\usepackage{lineno,hyperref}
\usepackage{graphicx}
\usepackage{amsmath}
\usepackage{color}
\usepackage[normalem]{ulem}
\usepackage{mathtools}

\usepackage{rotating}
\usepackage{lscape}

\journal{arXiv}
\begin{document}

\begin{frontmatter}

\title{Broken Symmetry of Stock Returns - a Modified Jones-Faddy Skew t-Distribution}

\author[mymainaddress]{Siqi Shao}
\author[mymainaddress]{Arshia Ghasemi}
\author[mymainaddress]{Hamed Farahani}
\author[mymainaddress]{R. A. Serota\fnref{myfootnote}}
\fntext[myfootnote]{serota@ucmail.uc.edu}

\address[mymainaddress]{Department of Physics, University of Cincinnati, Cincinnati, Ohio 45221-0011}

\begin{abstract}
We argue that negative skew and positive mean of the distribution of stock returns are largely due to the broken symmetry of stochastic volatility governing gains and losses. Starting with stochastic differential equations for stock returns and for stochastic volatility we argue that the distribution of stock returns can be effectively split in two - for gains and losses - assuming difference in parameters of their respective stochastic volatilities. A modified Jones-Faddy skew t-distribution utilized here allows to reflect this in a single organic distribution which tends to meaningfully capture this asymmetry. We illustrate its application on distribution of daily S\&P500 returns, including analysis of its tails.
\end{abstract}

\begin{keyword}
Student's  t-Distribution \sep Jones-Fadddy Skew t-Distribution \sep Stock Returns Gains and Losses \sep Stochastic Volatility \sep Power-Law Tails
\end{keyword}

\end{frontmatter}

\section{Introduction}

\begin{figure}[htbp]
    \centering
    \includegraphics[width=0.77\linewidth]{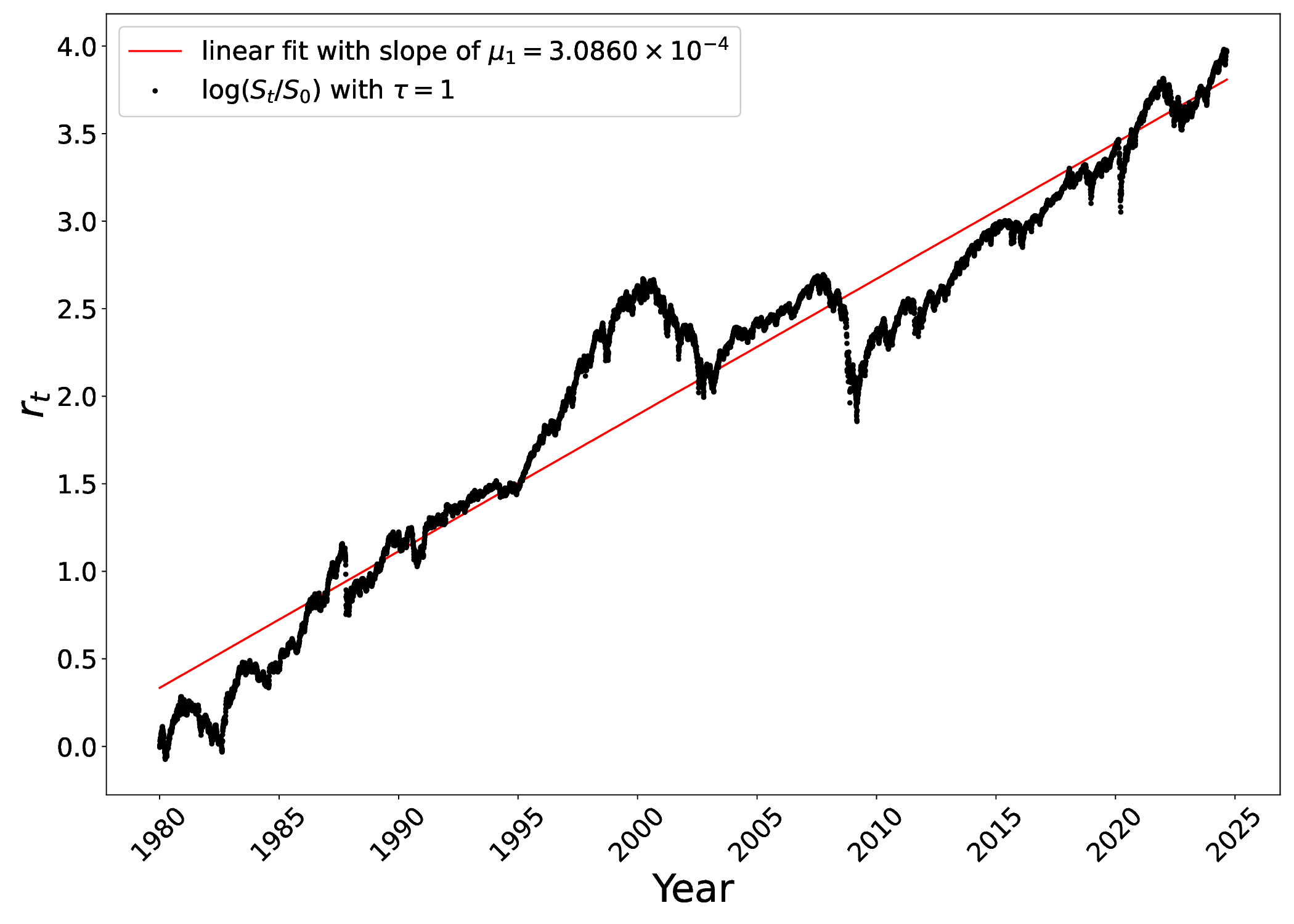}
    \caption{S\&P500; $r_t=\log(S_t/S_0)$, $S_t$ is price on day $t$, $t$ changes in daily increments ($\tau =1$ in text).}
    \label{rt}
\end{figure}

\begin{figure}[htbp]
    \centering
    \includegraphics[width=0.77\linewidth]{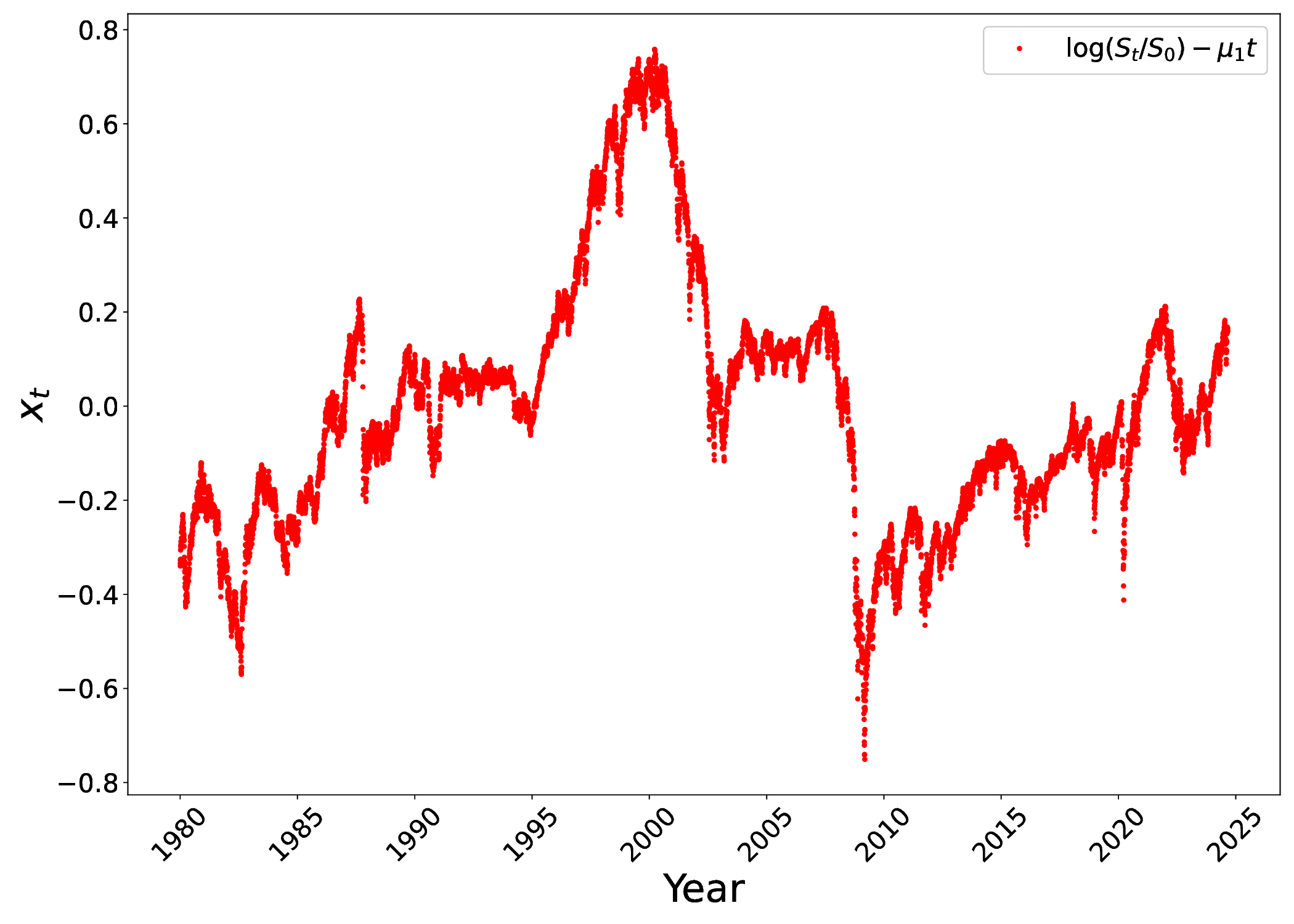}
    \caption{S\&P500; $x_t = r_t - \mu_1 t$ where index in $\mu_1$ reflects daily increments of $t$ ($\tau =1$ in text).}
    \label{xt}
\end{figure}

A well-known upward trend in stock prices is illustrated in Fig. \ref{rt} for S\&P500. Here $S_t$ is price on day $t$. The best linear fit $\mu t$ of $r_t = \log(S_t/S_0)$ corresponds to roughly $12\%$ annual growth. De-trended plot of returns $x_t = r_t - \mu t$ is shown in Fig. \ref{xt} and the fluctuations of $x_t$ are attributed to market volatility. The simplest of models attempting to describe these fluctuations implies that de-trended returns are governed by a stochastic differential equation (SDE)
\begin{equation}
dx_t =  \log\left(\frac{S_{t+\mathrm{d}t}}{S_t}\right) - \mu \mathrm{d}t = \sigma_t \mathrm{d}W^{(1)}
\label{dxt}
\end{equation}
where $\sigma_t$ is the stochastic volatility and $\mathrm{d}W = W\left(t+\mathrm{d}t\right) - W(t)$ is the normally distributed Wiener process, $\mathrm{d}W \sim \mathrm{N(}0,\, \mathrm{d}t \mathrm{)}$, $(\mathrm{d}W)^2 = \mathrm{d}t$. 

Stochastic volatility, in turn, is believed to be described by the mean-reverting SDE  for stochastic variance $v = \sigma_t^{2}$ 
\begin{equation}
\mathrm{d}v_t = -\gamma(v_t - \theta)\mathrm{d}t + g(v_t) \mathrm{d}W^{(2)}
\label{dvt}
\end{equation}
implying that stochastic variance - and hence volatility - tends to revert to its mean value, $\left<v_t\right>=\theta$. One of the important implications of the latter is that for returns accumulated over $\tau$ days, $\mathrm{d}t= \tau$, average variance of returns grows linearly with $\tau$, $\left<dx^{2}\right> = \theta \tau$. Since we are not concerned here with quantities such as leverage \cite{perello2003stochastic,dashti2021distributions} and study distribution of returns, in what follows we will neglect correlations between $\mathrm{d}W^{(1)}$ and $\mathrm{d}W^{(2)}$ \cite{dragulescu2002probability,liu2019distributions} and will largely concentrate on daily returns $\tau = 1$.

Numerous models exist for $g(v_t)$, such as Cox-Ingersoll-Ross \cite{cox1985theory,heston1993closed,dragulescu2002probability,liu2019distributions}, multiplicative \cite{praetz1972distribution,nelson1990arch,fuentes2009universal,liu2019distributions}, and the combination of the two \cite{dashti2021combined}. Here we will concentrate on multiplicative model since it is the simplest model that predicts power-law tails of the distribution of returns and is the easiest to handle analytically. While power-law tails in returns are not universally agreed upon, there is a strong case for them at least for daily returns, while for accumulated returns power law may persist for a large portion of the tail (see e.g. \cite{farahani2025asymmetry} and below).

In multiplicative model
\begin{equation} 
g(v_t)=\kappa v_t
\label{gvt}
\end{equation}
which yields an Inverse Gamma \cite{wolfram2025inverse} steady-state distribution (probability density function) for variance
\begin {equation}
f(v_t)=\mathrm{IGa(}v_t;\, \frac{\alpha}{\theta}+1,\, \alpha \mathrm{)}
\label{fvt}
\end {equation}
which, in turn, translates into the following distribution for volatility \cite{liu2019distributions}
\begin {equation}
f(\sigma_t)=2\sigma_t \cdot 
\mathrm{IGa(}\sigma_t^2;\, \frac{\alpha }{\theta}+1,\, \alpha \mathrm{)}
\label{fsigmat}
\end {equation}
where 
\begin {equation}
\alpha = \frac{2\gamma \theta}{\kappa^2}
\label{alpha}
\end{equation}
From eqs. (\ref{dxt}) and (\ref{fsigmat}) the distribution of stock returns can be found as a product distribution \cite{ma2014model} of inverse Gamma and normal distribution and the result is a Student t-distribution \cite{praetz1972distribution,fuentes2009universal,liu2019distributions}
\begin{equation}
f_{St}(x)=
\frac{\Gamma(\frac{\alpha}{\theta} + \frac{3}{2})}{\sqrt{\pi}\Gamma(\frac{\alpha}{\theta}+1)}\frac{1}{\sqrt{2\alpha \tau}}\left( \frac{x^2}{2\alpha \tau} + 1\right)^{-(\frac{\alpha}{\theta} + \frac{3}{2})}
\label{fSt}
\end{equation}

Clearly this distribution is even \footnote {In this particular case it is a consequence of normal distribution being even in the product distribution in (\ref{dxt}). It is also obvious that the product distribution will be even regardless of a specific form of $g(v_t))$} and thus establishes symmetry between gains and losses. This, of course, also applies to the power-law tails, whose exponent is $-\left(\frac{2\alpha}{\theta}+3\right)$. However this symmetry is clearly broken for actual data. To wit, the distribution of S\&P500 returns has \cite {farahani2025asymmetry}:
\begin{itemize}
	\item positive mean
	\item negative skew
	\item greater number of points for gains than for losses
	\item slower power-law exponent for losses than for gains
\end {itemize}

The motivation for this work was therefore to model this symmetry breaking while, ideally, still remaining within SDE framework. The first and fairly obvious idea would be that stochastic volatility eqs. (\ref{dvt}) and (\ref{gvt}) are governed by a different set of parameters for gains and losses, that is to say that their parameters $\alpha$ and $\theta$ are different. In other words, this implies that gains and losses should be fitted separately by weighted Student t-distributions (\ref{fSt}) in a manner that weights add up to unity and their ratio is the ratio of points under respective distributions. For brevity, we call the final distribution ``half Student-t''. 

While still having SDE underpinning ``half Student-t'' is obviously not an organic distribution. Additionally, it predicts a negative mean contrary to the empirical evidence above. Consequently, we adopted yet another approach based on a skew extension of Student t-distribution by Jones and Faddy. \cite{jones2001skew, jones2003skew}. Unfortunately, modified Jones-Faddy (mJF) distributions that we use are not cleanly derived from SDE formalism but on the other hand they are close in spirit and yield good fits to the full distribution of returns. 

This paper is organized as follows. In Section \ref{AF} we provide expressions for the probability density function (PDF) and cumulative distribution function (CDF) of mJF distributions as well as their statistical parameters, such as mean, mode, variance and skewness. In Section \ref{NR} we present results of fitting the full distribution of returns. Finally, we summarize and discuss our results in Section \ref{CD}.

\section{Analytical Framework \label{AF}}
In this section we give analytical expressions for PDF, CDF, mean $m_1$, variance $m_2$ ($m_2^{1/2}$ being standard deviation), and mode $\overline{m}$ for four distributions described in this section. We use first and second Pearson coefficients of skewness
\begin{equation}
\zeta_1=\frac{\left(m_1-\overline{m}\right)}{m_2^{1/2}}, \qquad \zeta_2=\frac{\left(m_1-\widetilde{m}\right)}{m_2^{1/2}}
\label{zeta12}
\end{equation}
to characterize the skew of the distribution, where $\widetilde{m}$ is the median, which is evaluated numerically. \footnote{See Appendix for an alternative derivation of the median.} \footnote{The third moment and Fisher-Pearson coefficient diverge due to slow power-law decays of the tails per Table \ref{stats} and \cite{farahani2025asymmetry}.} We will consider CDF appropriate for gains and losses separately per, respectively,
\begin{equation}
	F_g (x) = \int_{-\infty}^{x} f(y) \mathrm{d}y \quad\text{and}\quad 	F_l (x) = \int_{x}^{\infty} f(y) \mathrm{d}y 
\label{FgFl}
\end{equation}
where $f(x)$ is the PDF of returns. Complementary CDF, CCDF, appropriate for gains and losses are, respectively, $1-F_g (x)$ and $1-F_l (x)$.

\subsection{Student t-Distribution\label{St}}
PDF of Student t-distribution is given by (\ref{fSt}) implying that the tails of PDF scale as $x^{-\left( 2 \alpha/\theta+3 \right)}$ and of CDF as $x^{-2 \left( \alpha/\theta+1 \right)}$. Due to symmetry CDF for both gains and losses is given by 
\begin{equation}
F_{St} (x) = \frac{1}{2} \left(1 +  I\!\left(
\frac{x^2}{x^2 + 2 \alpha \tau}; \,
\frac{1}{2}, \,
1 + \frac{\alpha}{\theta}
\right)\right)
\label{FSt}
\end{equation}
where $I(x;a,b)$ is a regularized incomplete beta function \cite{nist2025digital} and, obviously, $m_1=\overline{m}=\widetilde{m}=\zeta_{1,2}=0$ and
\begin{equation}
m_2=\theta \tau
\label{m2St}
\end{equation}

\subsection{Half Student-t Distribution\label{hSt}}
In effect, it is a mixture distribution whose PDF of gains and losses are given, respectively, by 
\begin{equation}
f_{g\_hSt} (x) = \frac{2 \, \Gamma\!\left(\frac{\alpha_g}{\theta_g} + \frac{3}{2}\right)}
{\sqrt{\pi}\,\Gamma\!\left(\frac{\alpha_g}{\theta_g} + 1\right)}
\frac{1}{\sqrt{2\alpha_g \tau}}
\left(
\frac{x^{2}}{2\alpha_g \tau} + 1
\right)^{-\left(\frac{\alpha_g}{\theta_g} + \frac{3}{2}\right)}, \qquad x\ge 0
\label{fhStg}
\end{equation}
and
\begin{equation}
f_{l\_hSt} (x) =\frac{2 \, \Gamma\!\left(\frac{\alpha_l}{\theta_l} + \frac{3}{2}\right)}
{\sqrt{\pi}\,\Gamma\!\left(\frac{\alpha_l}{\theta_l} + 1\right)}
\frac{1}{\sqrt{2\alpha_l \tau}}
\left(
\frac{x^{2}}{2\alpha_l \tau} + 1
\right)^{-\left(\frac{\alpha_l}{\theta_l} + \frac{3}{2}\right)}, \qquad x\le 0
\label{fhStl}
\end{equation}
so that the full distribution is given by 
\begin{equation}
f_{hSt}= w_g f_{g\_hSt} + w_l f_{l\_hSt} (x)
\label{fhSt}
\end{equation}
where $w_g+w_l=1$ and $w_g/w_l$ is the ratio of points under gains and losses \cite{farahani2025asymmetry}. Generally speaking a mixture distribution is not a preferable venue from a physicist's point of view since it does not follow form a first--principles model. However $f_{g\_hSt} $ and $f_{l\_hSt}$ marginally do.

CDF of gains and losses are given, respectively, by 
\begin{equation}
F_{g\_hSt}(x) = w_l + w_g I\!\left(
\frac{x^2}{x^2 + 2 \alpha_g \tau}; \,
\frac{1}{2}, \,
1 + \frac{\alpha_g}{\theta_g}
\right)
\label{FhStg}
\end{equation}
and 
\begin{equation}
F_{l\_hSt}(x)  = w_g + w_l I\!\left(
\frac{x^2}{x^2 + 2 \alpha_l \tau}; \,
\frac{1}{2}, \,
1 + \frac{\alpha_l}{\theta_l}
\right)
\label{FhStl}
\end{equation}

Using (\ref{fhStg})-(\ref{fhSt}) we find the following expressions for the mean and variance respectively
\begin{equation}
m_1 = \sqrt{\frac{2}{\pi}}
\left(
\frac{
w_g \sqrt{\alpha_g \tau}\,\Gamma\!\left(\tfrac{1}{2} + \tfrac{\alpha_g}{\theta_g}\right)
}{
\Gamma\!\left(1+\tfrac{\alpha_g} {\theta_g}\right)
}
-
\frac{
w_l \sqrt{\alpha_l \tau}\,\Gamma\!\left(\tfrac{1}{2} + \tfrac{\alpha_l}{\theta_l}\right)
}{
\Gamma\!\left(1+\tfrac{\alpha_l} {\theta_l}\right)
}
\right)
\label{m1hSt}
\end{equation}
and
\begin{align}
m_2 &= \tau \Bigg(
w_g \theta_g + w_l \theta_l \notag\\
&\quad + \frac{
2(-2 + w_g + w_l) \Big(
\scriptstyle w_l^2 \alpha_l \Gamma\!\left(1+\frac{\alpha_g} {\theta_g}\right)^2
\Gamma\!\left(\frac{1}{2} + \frac{\alpha_l}{\theta_l}\right)^2
- 2 w_g w_l \sqrt{\alpha_g \alpha_l} 
\Gamma\!\left(\frac{1}{2} + \frac{\alpha_g}{\theta_g}\right)
\Gamma\!\left(1+\frac{\alpha_g} {\theta_g}\right)
\Gamma\!\left(\frac{1}{2} + \frac{\alpha_l}{\theta_l}\right)
\Gamma\!\left(1+\frac{\alpha_l} {\theta_l}\right)
+ w_g^2 \alpha_g 
\Gamma\!\left(\frac{1}{2} + \frac{\alpha_g}{\theta_g}\right)^2
\Gamma\!\left(1+\frac{\alpha_l} {\theta_l}\right)^2
\Big)
}{
\scriptstyle \pi \Gamma\!\left(1+\frac{\alpha_g} {\theta_g}\right)^2
\Gamma\!\left(1+\frac{\alpha_l} {\theta_l}\right)^2
}
\Bigg)
\label{m2hSt}
\end{align}
where $\Gamma(x)$ is a Gamma function \cite{nist2025digital}. Clearly, $\overline{m}=0$ in this model so the sign of the skew $\zeta_1$ will be that of $m_1$. Another observation is that the number of parameters in half Student-t, aside from $\tau$, is double that of Student t-distribution. One possible simplification of this model is to assume that the mean stochastic volatility governing gains and losses is the same, $\theta_g = \theta_l = \theta$ so that the difference between gains and losses, including power-law exponents, reduces solely to difference between $\alpha_g$ and $\alpha_l$.

\subsection{Modified Jones-Faddy Distribution mJF1\label{mJF1}}
PDF of the first of modified Jones-Faddy distributions (mJF1) introduced here for characterization of distribution of stock returns is given by
\begin{equation}
f(x)=C\left(1-\frac{x-\mu}{\sqrt{(x-\mu)^2+(\alpha_g+\alpha_l) \tau}}\right)^{\frac{\alpha_g}{\theta}+\frac{3}{2}} \left(1+\frac{x-\mu}{\sqrt{(x-\mu)^2+(\alpha_g+\alpha_l) \tau}}\right)^{\frac{\alpha_l}{\theta}+\frac{3}{2}}
\label{fmJF1}
\end{equation}
where the normalization factor $C$ is given by
\begin{equation}
C=\frac{1}{2^{\frac{\alpha_l}{\theta}+1+\frac{\alpha_g}{\theta}} \text{B}(\frac{\alpha_l}{\theta}+1,\frac{\alpha_g}{\theta}+1)  } \frac{1}{\sqrt{(\alpha_g+\alpha_l)\tau}}
\label{CmJF1}
\end{equation}
CDF for gains and losses are given, respectively, by 
\begin{equation}
F_{g\_mJF1} (x) =I\left(1+\frac{x-\mu}{\sqrt{(x-\mu)^2+(\alpha_g+\alpha_l) \tau}}; \frac{\alpha_g}{\theta} + 1,\frac{\alpha_l}{\theta} + 1 \right)
\label{FmJF1g}
\end{equation}
and
\begin{equation}
F_{l\_mJF1} (x) =I\left(1-\frac{x-\mu}{\sqrt{(x-\mu)^2+(\alpha_g+\alpha_l) \tau}}; \frac{\alpha_g}{\theta} + 1,\frac{\alpha_l}{\theta} + 1\right)
\label{FmJF1l}
\end{equation}
mJF1 is a direct descendent of the distribution (\ref{fSt}) with one minor and one significant variation. First, as in standard Student distribution \cite{wolfram2025student}, a location parameter $\mu$ can be introduced and is introduced here. Obviously it does not affect (\ref{dxt}) since the variable can always be shifted by a constant. The second variation introduces a skew (skew t distribution \cite{jones2001skew,jones2003skew}), via $\alpha_g$ and $\alpha_l$ here. In particular, power-law tails scale as $x^{-\left( 2 \alpha_g/\theta+3 \right)}$ at $+\infty$ and $x^{-\left( 2 \alpha_l/\theta+3 \right)}$ at $-\infty$. This brakes a construct based on (\ref{dxt}) and (\ref{dvt}) which treats volatility of gains and losses uniformly: substituition $\alpha_g=\alpha_l=\alpha$ in (\ref{fmJF1}) leads back to (\ref{fSt}) (with non-zero location parameter $\mu$). At this point we are unaware of an SDE-based formulation which would result in a distribution (\ref{fmJF1}).

Turning now to mean, variance and mode of mJF1 we find, respectively,

\begin{equation}
m_1 =\mu +\sqrt{(\alpha_g+\alpha_l)\tau}\, 
	\mathrm{B}\!\left(\frac{\alpha_g}{\theta}+\frac{1}{2},\frac{1}{2}\right)\,
       \mathrm{B}\!\left(\frac{\alpha_l}{\theta}+\frac{1}{2},\frac{1}{2}\right)\,
        \frac{\dfrac{\alpha_l}{\theta}-\dfrac{\alpha_g}{\theta}}{2\pi}\\
\label{m1mJF1}
\end{equation}
\begin{equation}
m_2 = \theta\tau
\frac{(\alpha_g+\alpha_l)^2}{4\alpha_g\alpha_l}
+\frac{(\alpha_g+\alpha_l)(\alpha_g-\alpha_l)^2\tau}{4\theta^2}
\left[
\frac{\theta^2}{\alpha_g\alpha_l}
-\left(\frac{\pi}{
\mathrm{B}\!\left(\frac{\alpha_g}{\theta},\frac{1}{2}\right)
\mathrm{B}\!\left(\frac{\alpha_l}{\theta},\frac{1}{2}\right)}\right)^2
\right].
\label{m2mJF1}
\end{equation}
\begin{equation}
\overline{m}=\mu + \sqrt{(\alpha_g+\alpha_l)\tau} \frac{\dfrac{\alpha_l}{\theta}-\dfrac{\alpha_g}{\theta}}
{2\sqrt{\left(\dfrac{\alpha_g}{\theta}+\dfrac{3}{2}\right)
\left(\dfrac{\alpha_l}{\theta}+\dfrac{3}{2}\right)}},
\label{mbarmJF1}
\end{equation}

\subsection{Modified Jones-Faddy Distribution mJF2 \label{mJF2}}

The second modified Jones-Faddy distribution mJF2 is a simple generalization of mJF1 in that instead of a single $\theta$ we now have $\theta_g$ and $\theta_l$, just as for half Student-t in Sec. \ref{hSt}. At this point we believe that assumption of the same mean stochastic volatility for gains and losses, as is the case for mJF1, makes more sense. Additionally, introduction of an extra fitting parameter in mJF2 only minimally improves fitting. Therefore, we present mJF2 results largely for completeness.

PDF of mJF2 is given by
\begin{equation}
f(x)=C\left(1-\frac{x-\mu}{\sqrt{(x-\mu)^2+(\alpha_g+\alpha_l) \tau}}\right)^{\frac{\alpha_g}{\theta_g}+\frac{3}{2}} \left(1+\frac{x-\mu}{\sqrt{(x-\mu)^2+(\alpha_g+\alpha_l) \tau}}\right)^{\frac{\alpha_l}{\theta_l}+\frac{3}{2}}
\label{fmJF2}
\end{equation}
where normalization factor is
\begin{equation}
C=\frac{1}{2^{\frac{\alpha_l}{\theta_l}+1+\frac{\alpha_g}{\theta_g}} \text{B}(\frac{\alpha_l}{\theta_l}+1,\frac{\alpha_g}{\theta_g}+1)  } \frac{1}{\sqrt{(\alpha_g+\alpha_l)\tau}}
\label{CmJF2}
\end{equation}
CDF for gains and losses are given, respectively, by 
\begin{equation}
F_{g\_mJF2} (x) =I\left(1+\frac{x-\mu}{\sqrt{(x-\mu)^2+(\alpha_g+\alpha_l) \tau}}; \frac{\alpha_g}{\theta_g} + 1,\frac{\alpha_l}{\theta_l} + 1 \right)
\label{FmJF2g}
\end{equation}
and
\begin{equation}
F_{l\_mJF2} (x) =I\left(1-\frac{x-\mu}{\sqrt{(x-\mu)^2+(\alpha_g+\alpha_l) \tau}}; \frac{\alpha_g}{\theta_g} + 1,\frac{\alpha_l}{\theta_l} + 1\right)
\label{FmJF2l}
\end{equation}

Mean, variance and mode of mJF2 are given, respectively, by
\begin{equation}
m_1 =\mu +\sqrt{(\alpha_g+\alpha_l)\tau}\,
	\mathrm{B}\!\left(\frac{\alpha_g}{\theta_g}+\frac{1}{2},\frac{1}{2}\right)\,
	\mathrm{B}\!\left(\frac{\alpha_l}{\theta_l}+\frac{1}{2},\frac{1}{2}\right)\,
	\frac{\dfrac{\alpha_l}{\theta_l}-\dfrac{\alpha_g}{\theta_g}}{2\pi} \\
\label{m1mJF2}
\end{equation}
\begin{equation}
m_2
= (\theta_l\alpha_g+\theta_g\alpha_l)\tau
\frac{\alpha_g+\alpha_l}{4\alpha_g\alpha_l}
+\frac{(\alpha_g+\alpha_l)\tau}{4}\left(\frac{\alpha_g}{\theta_g}-\frac{\alpha_l}{\theta_l}\right)^2
\left[
\frac{\theta_g\theta_l}{\alpha_g\alpha_l}
-\left(\frac{\pi}{
\mathrm{B}\!\left(\frac{\alpha_g}{\theta_g},\frac{1}{2}\right)
\mathrm{B}\!\left(\frac{\alpha_l}{\theta_l},\frac{1}{2}\right)}\right)^2
\right]
\label{m2mJF2}
\end{equation}
\begin{equation}
\overline{m}
= \mu + \sqrt{(\alpha_g+\alpha_l)\tau}\,
\frac{\dfrac{\alpha_l}{\theta_l} - \dfrac{\alpha_g}{\theta_g}}
{2\sqrt{\left(\dfrac{\alpha_g}{\theta_g}+\dfrac{3}{2}\right)
\left(\dfrac{\alpha_l}{\theta_l}+\dfrac{3}{2}\right)}}.
\label{mbarmJF2}
\end{equation}

\newpage

\section{Numerical Results \label{NR}}
Table \ref{params} shows parameters of distributions in Sec. \ref{AF} obtained by Bayesian fitting of  1980-2025 S\&P500 returns. Table \ref{stats} gives the values of mean $m_1$, variance $m_2$, and mode $\overline{m}$ from equations obtained in that section for each of the distributions. $\widetilde{m}$ is evaluated numerically. First and second Pearson coefficients of skewness, $\zeta_1$ and $\zeta_2$ are then computed using (\ref{zeta12}). Exponents of power-law tails of the distributions' CCDF are computed as $-2 \left(\frac{\alpha_i}{\theta_i}+1\right)$, where $\alpha_i = \alpha, \alpha_g, \alpha_l$ and  $\theta_i = \theta, \theta_g, \theta_l$. Tail exponents of S\&P500 returns are obtained by linear fitting of the tails.
\begin{table}[htbp]

   \centering

   \scriptsize

   \begin{tabular}{|c|c|c|c|c|c|c|c|}
       \hline
       Simulations & $\theta$ & $\theta_{g}$ & $\theta_{l}$ & $\alpha$ &$\alpha_{g}$ & $\alpha_{l}$ & $\mu$ \\
       \hline
       Student-t&$1.407 \times 10^{-4}$& - & - & $7.347 \times 10^{-5}$&-&-&-  \\
       \hline
        Half Student-t & - &$1.182 \times 10^{-4}$ &$1.803 \times 10^{-4}$ &- & $8.512 \times 10^{-5}$&$6.134 \times 10^{-5}$&- \\
        \hline
        mJF1&$1.422\times 10^{-4}$ & -& -& - &$ 7.924\times 10^{-5}$  & $6.416\times 10^{-5}$ &$8.465\times 10^{-4}$ \\
       \hline
       mJF2 & - &$1.197 \times 10^{-4}$ &$1.778 \times 10^{-4}$&- & $6.393 \times 10^{-5}$&$7.634 \times 10^{-5}$& $8.623\times 10^{-4}$ \\
       \hline
   \end{tabular}
   \caption{Fitting parameters of distributions (\ref{fSt}), (\ref{fhSt}), (\ref{fmJF1}), and (\ref{fmJF2}).}
   \label{params}
\end{table}

\begin{table}[htbp]
   \centering
  \scriptsize
   \begin{tabular}{|c|c|c|c|c|c|c|clcl}
       \hline
       Simulations & $m_{1}$ & $m_{2}$ & $\overline{m}$  & $\zeta_{1}$ &$ \widetilde{m}$ & $\zeta_{2}$& Gains & Losses\\
       \hline
       Student-t&0&$1.41 \times 10^{-4}$& 0 & 0 & 0 &0 &-3.04 &-3.04 \\
       \hline
        Half Student-t &$-2.47 \times 10^{-4} $ & $1.48 \times 10^{-4} $ & 0& $-2.03\times 10^{-2}$ &$6.047 \times 10^{-6}$ &$2.08 \times 10^{-2}$ &-3.04 &-2.95\\
        \hline
        mJF1&$4.06 \times 10^{-5}$ & $1.44 \times 10^{-4} $ & $5.22 \times 10^{-4}$ &$-3.96\times 10^{-2}$  & $3.211 \times 10^{-4}$& $2.29 \times 10^{-2}$ &-3.12 &-2.90\\
       \hline
       mJF2 &$5.29\times 10^{-5}$ & $1.45 \times 10^{-4}$ & $5.49 \times 10^{-4}$ & $-4.49\times 10^{-2}$& $3.395 \times 10^{-4}$& $2.65 \times 10^{-2}$ &-3.07&-2.76 \\
       \hline
       S\&P500 &$4.38\times 10^{-5}$ & $1.28\times 10^{-4}$ & $1.32\times 10^{-4}$ & $-7.70 \times 10^{-3}$ &$2.733 \times 10^{-4}$ & $2.03 \times 10^{-2}$ & -3.14 & -2.91 \\
       \hline
    \end{tabular}
   \caption{\centering Mean, variance, mode, first Fisher skewness coefficient, median, second Fisher skewness coefficient, and exponents of power-law tails of distributions in Table \ref{params} and S\&P500 returns.}
   \label{stats}
\end{table}

\begin{figure}[htbp]
    \centering
    \includegraphics[width=1\linewidth]{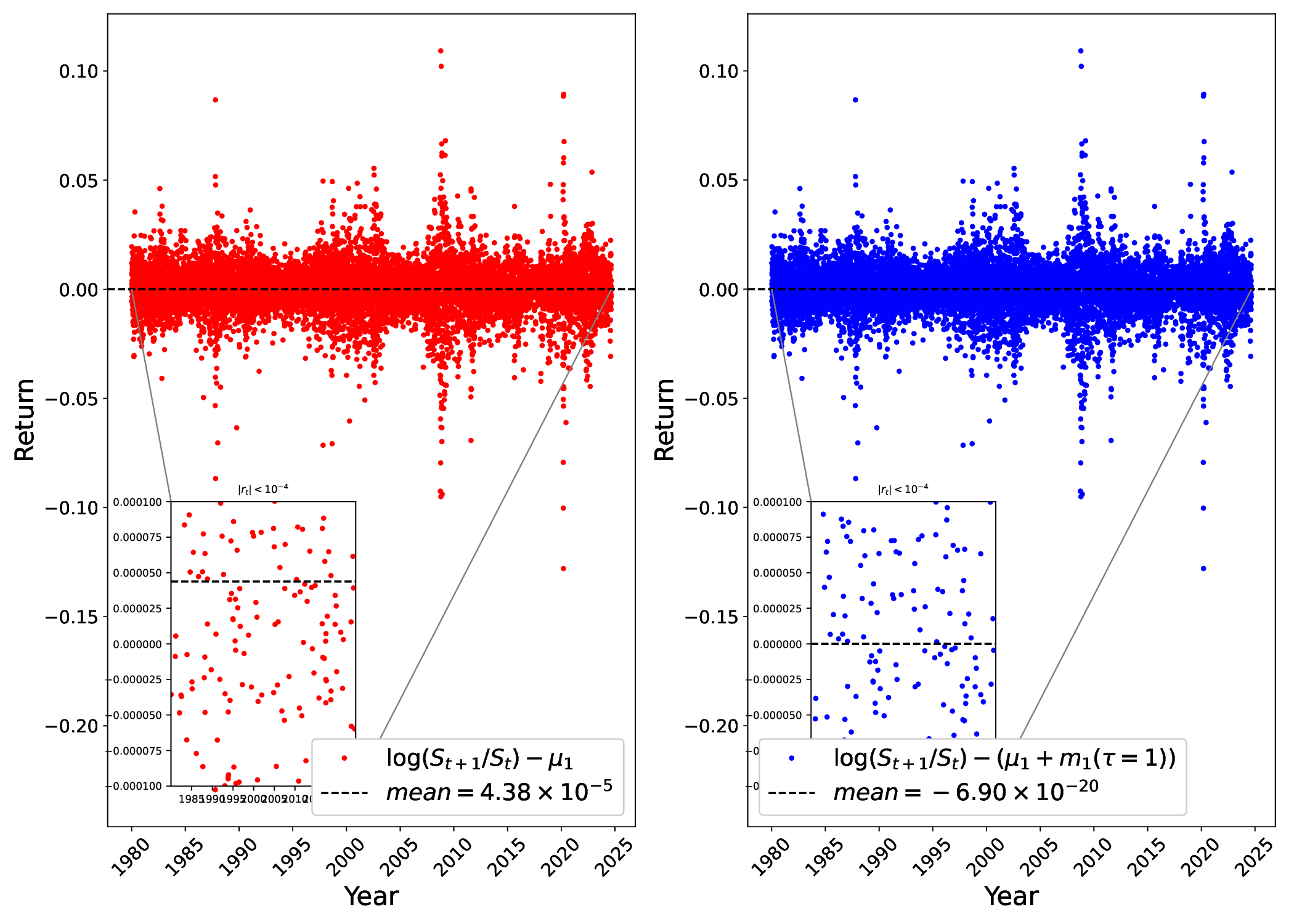}
    \caption{left: $x_t$ (\ref{xt}) for S\&P daily returns ($\tau=1$) -- centered on $m_1$; right: $x_t-m_1$($\tau=1$) -- centered on $0$.}
    \label{xtm1}
\end{figure}

Clearly, half Student-t, which does not allow for location parameter, fails to capture positive sign of $m_1$. We point out that positive values of $m_1$ in Table \ref{stats} are roughly an order of magnitude smaller than $\mu_1$ of the linear fit in Fig. \ref{rt} but are still noon-zero, as illustrated in Fig. \ref{xtm1}. Also, the second Fisher coefficients of skewness $\zeta_2$ of fitted distribution are much closer to S\&P500 returns than the first Fisher coefficients $\zeta_1$. Specifically for mJF1 and mJF2 it is due to the large discrepancy in the value of the mode $\overline{m}$. The reason for that is difficulty in identifying the value of the mode in empirical data, as is obvious from Fig. \ref{fplotmax} below (see next paragraph for explanation). In this particular case we used a smoothing procedure to obtain the value of $\overline{m}$ for S\&P500.

\begin{figure}[htbp]
    \centering
    \includegraphics[width=.77\linewidth]{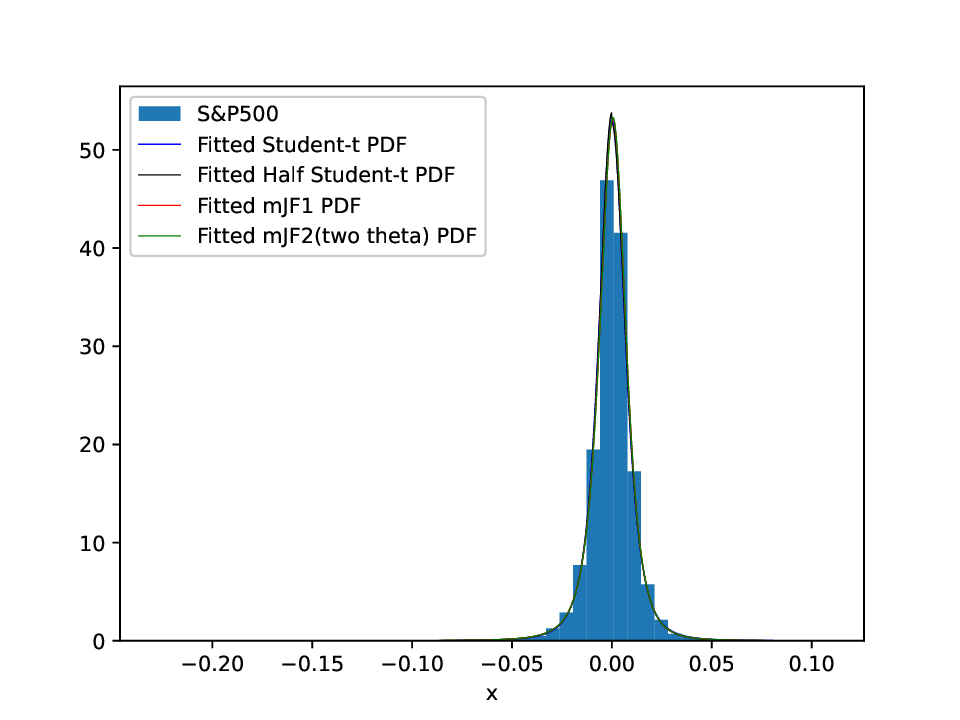}
    \caption{PDF of stocks returns and fits with distributions (\ref{fSt}), (\ref{fhSt}), (\ref{fmJF1}), and (\ref{fmJF2}).}
    \label{fplot}
\end{figure}

\begin{figure}[htbp]
    \centering
    \includegraphics[width=0.77\linewidth]{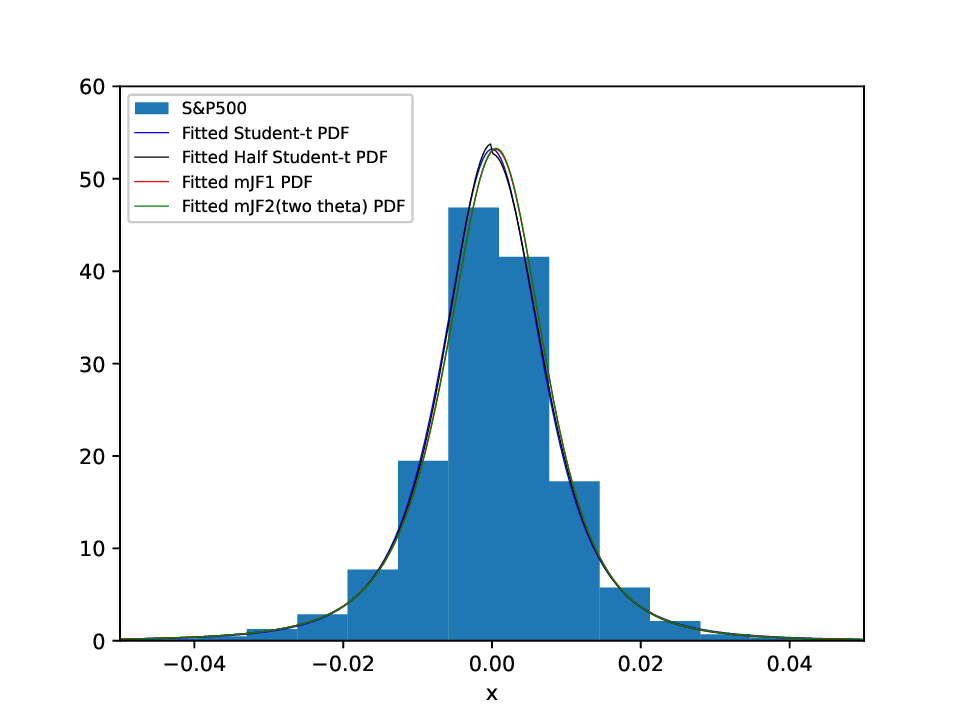}
    \caption{Expanded view of the area around the mode of the distribution in Fig. \ref{fplot}.}
    \label{fplotmax}
\end{figure}

\begin{figure}[htbp]
    \centering
    \includegraphics[width=0.77\linewidth]{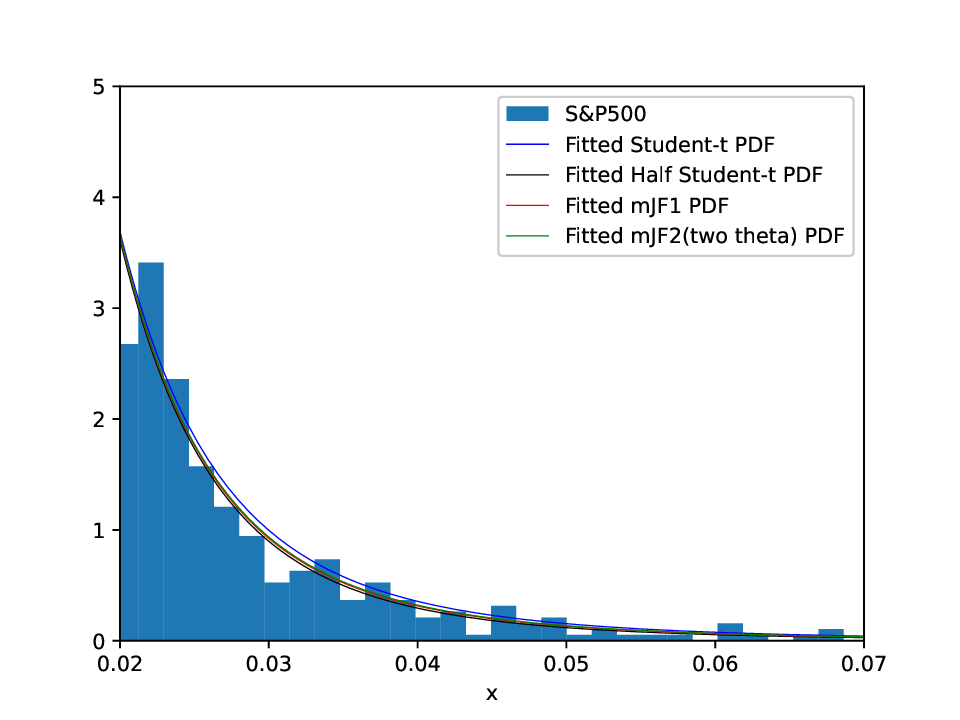}
    \caption{Expanded view of the tail area of Fig. \ref{fplot} for gains.}
    \label{fplotg}
\end{figure}

\begin{figure}[htbp]
    \centering
    \includegraphics[width=0.77\linewidth]{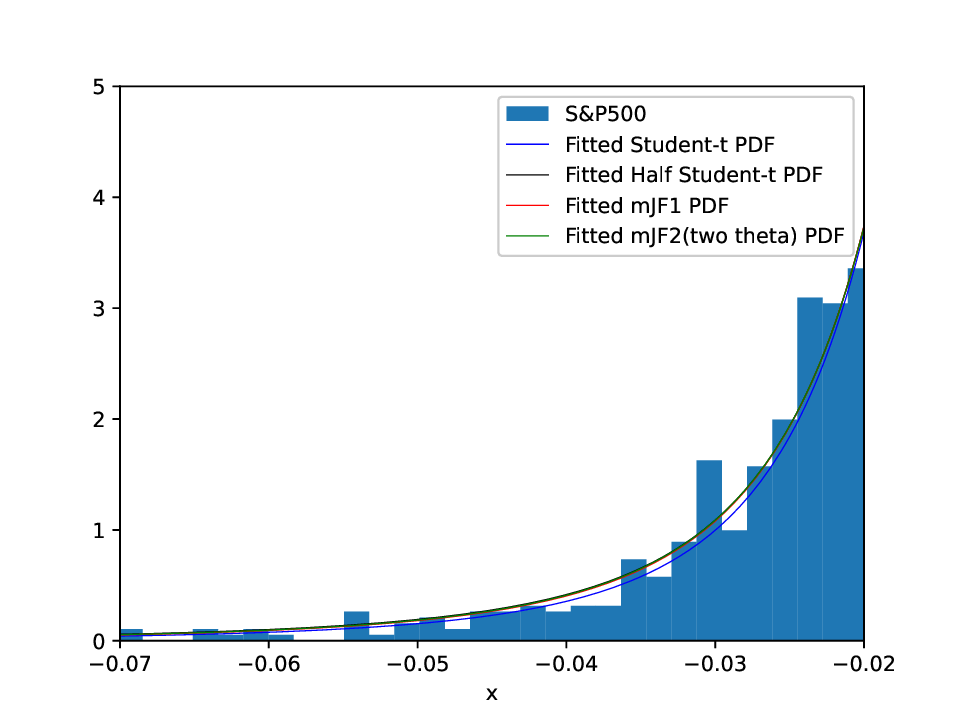}
    \caption{Expanded view of the tail area of Fig. \ref{fplot} for losses.}
    \label{fplotl}
\end{figure}

\begin{figure}[htbp]
    \centering
    \includegraphics[width=0.77\linewidth]{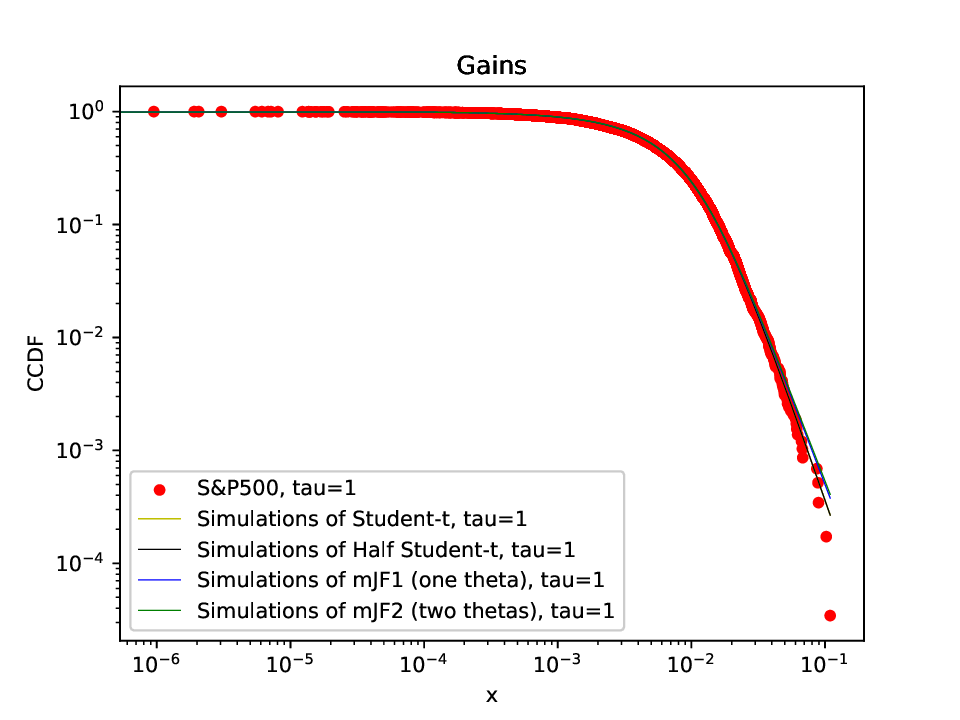}
    \caption{CCDF of gains with CCDF of fitting distributions.}
    \label{Fplotg}
\end{figure}

\begin{figure}[htbp]
    \centering
    \includegraphics[width=0.77\linewidth]{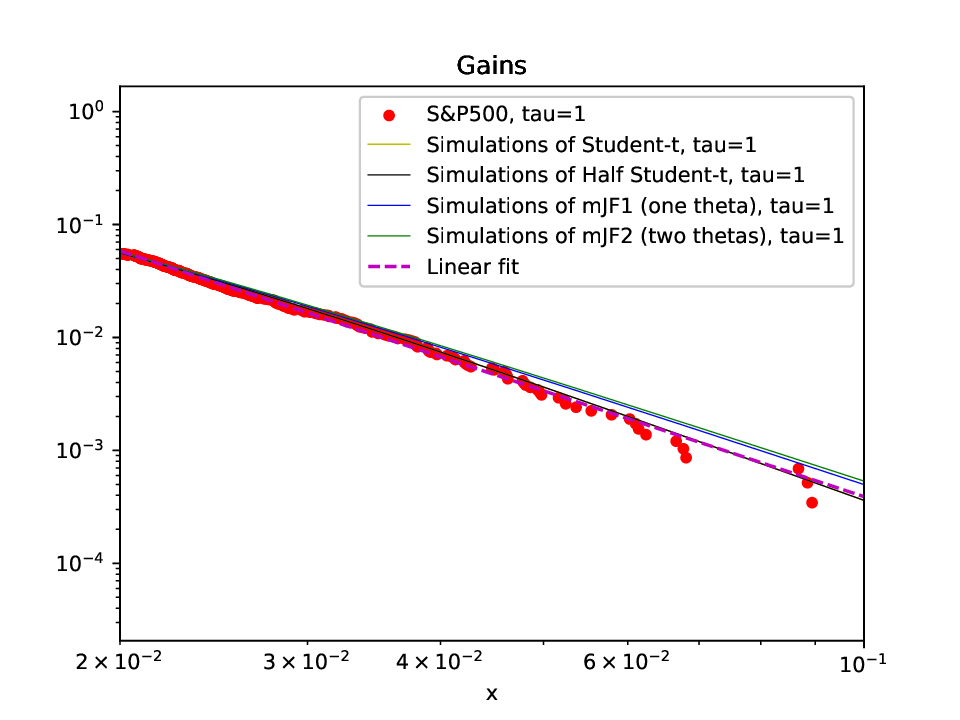}
    \caption{Expanded view of the tail area of Fig. \ref{Fplotg}.}
    \label{Fplottailg}
\end{figure}

\begin{figure}[htbp]
    \centering
    \includegraphics[width=0.77\linewidth]{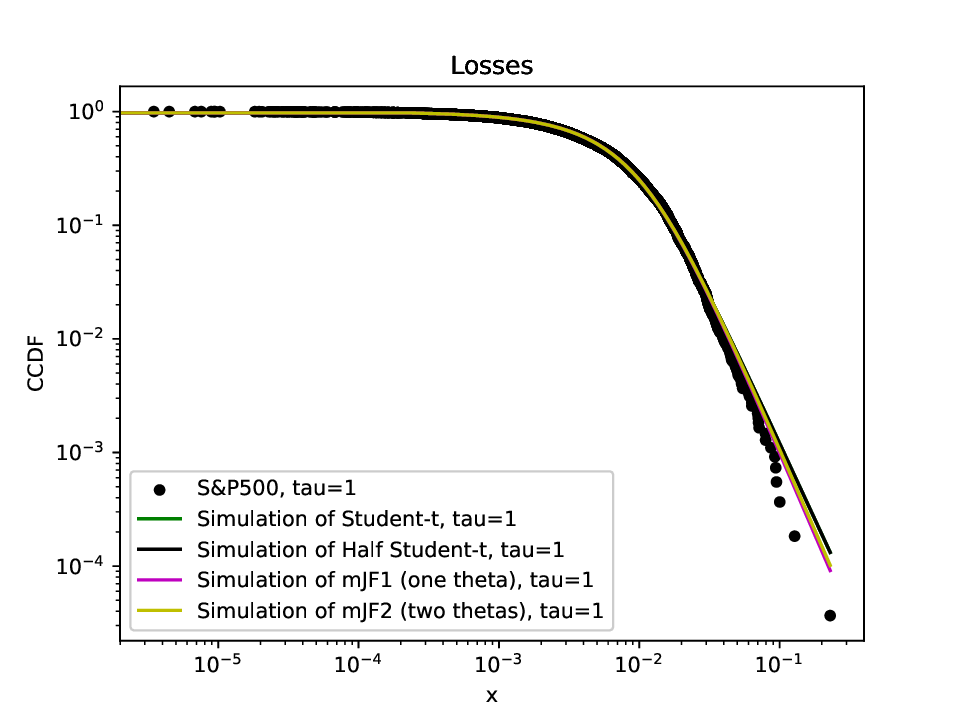}
    \caption{CCDF of losses with CCDF of fitting distributions.}
    \label{Fplotl}
\end{figure}

\begin{figure}[htbp]
    \centering
    \includegraphics[width=0.77\linewidth]{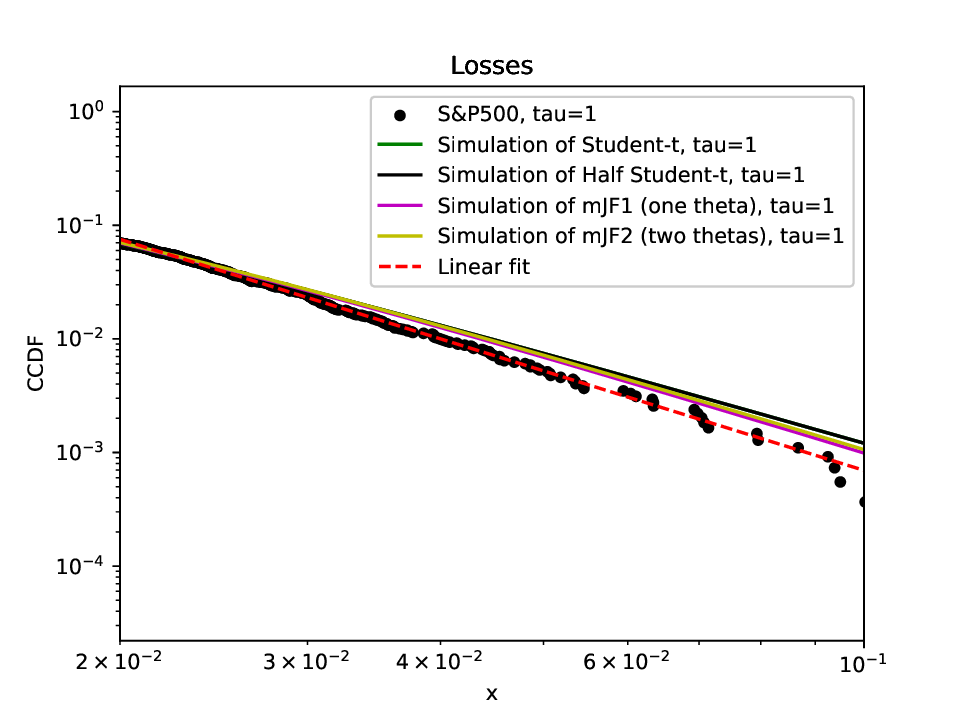}
    \caption{Expanded view of the tail area of Fig. \ref{Fplotl}.}
    \label{Fplottaill}
\end{figure}

Figs. \ref{fplot}-\ref{fplotl} show fits of the PDF of the distribution of S\&P500 returns using PDF of four distributions described in Sec. \ref{AF}. These fits render parameters listed in Table \ref{params}. Figs. \ref{Fplotg}-\ref{Fplottaill} show CCDF of S\&P500 returns and CCDF of the four fitting distributions derived  in Sec. \ref{AF} with parameters from Table \ref{params}. Linear fits of tail areas are also shown in Figs. \ref{Fplottailg} and \ref{Fplottaill}. Clearly, visually all four distributions exhibit very close  tail behavior, which also only slightly differs from S\&P500 tail and its linear fit. This is confirmed by closeness of power-law exponents in Table \ref{stats}. We point out, however, that the distributions of Sec. \ref{AF} approach power-law behavior only asymptotically and their own linear fits in Figs. \ref{Fplottailg} and \ref{Fplottaill}would not generally speaking produce exponents listed in Table \ref{AF}.

\begin{figure}[htbp]
    \centering
    \includegraphics[width=0.7\linewidth]{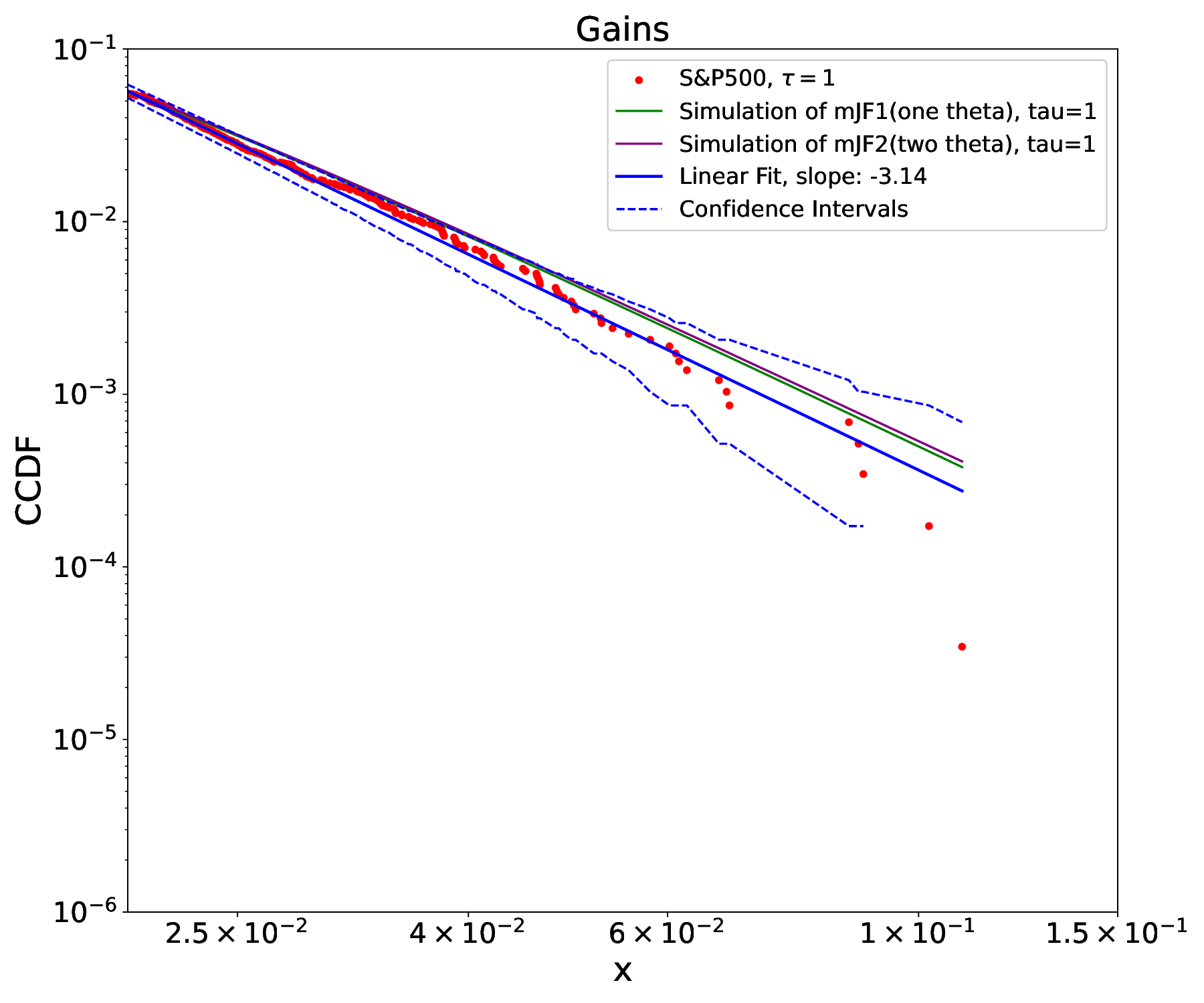}
   \caption{ \centering
Linear fit, with its confidence interval, of the tail of the S\&P500 distribution of daily gains; mJF1 and mJF2 fits are shown for comparison.}
    \label{LFCIg}
\end{figure}

\begin{figure}[htbp]
    \centering
    \includegraphics[width=0.7\linewidth]{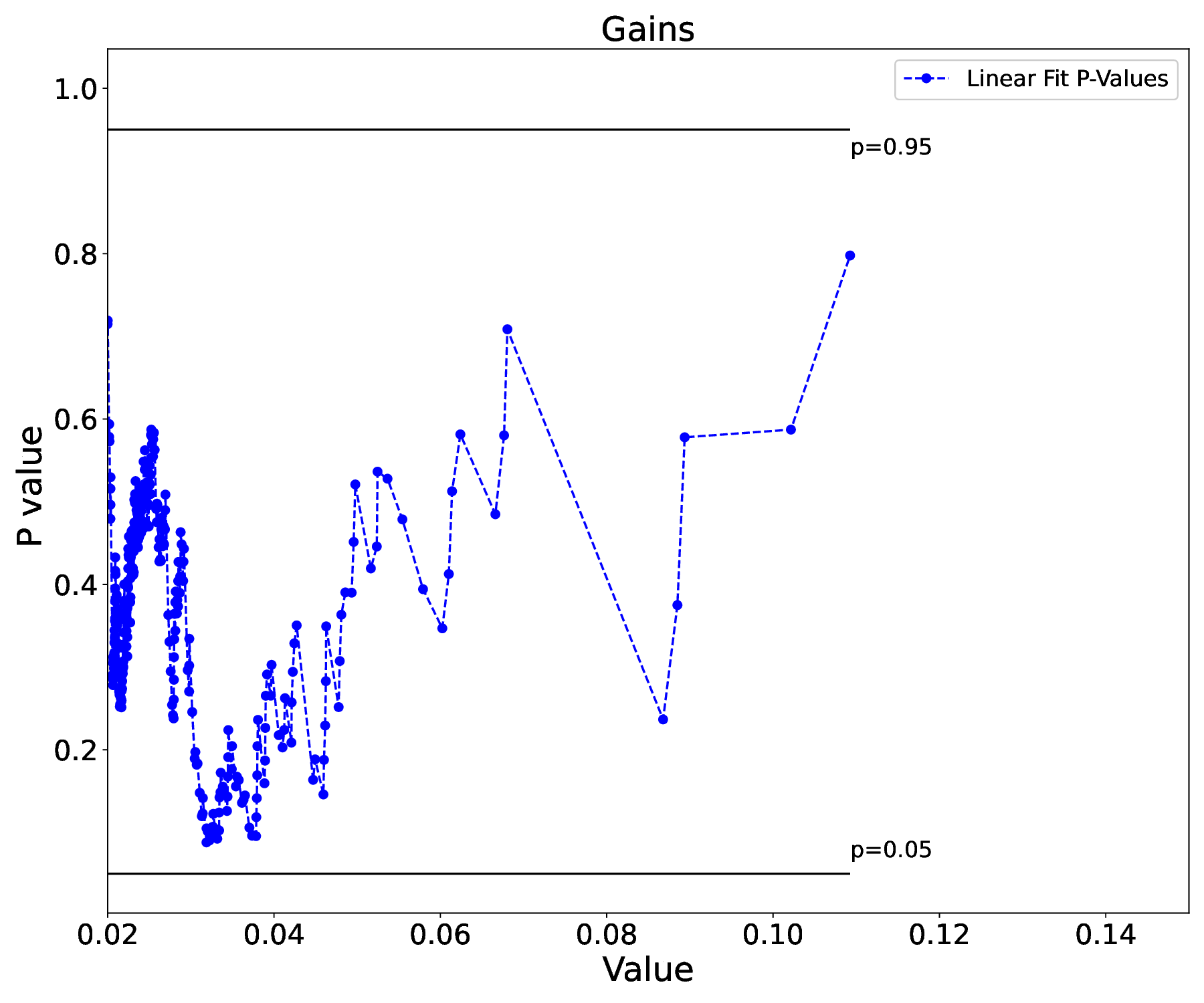}
    \caption{p-values of statistical U-test for the linear fit of the S\&P500 distribution of daily gains.}
    \label{LFPVg}
\end{figure}

\begin{figure}[htbp]
    \centering
    \includegraphics[width=0.7\linewidth]{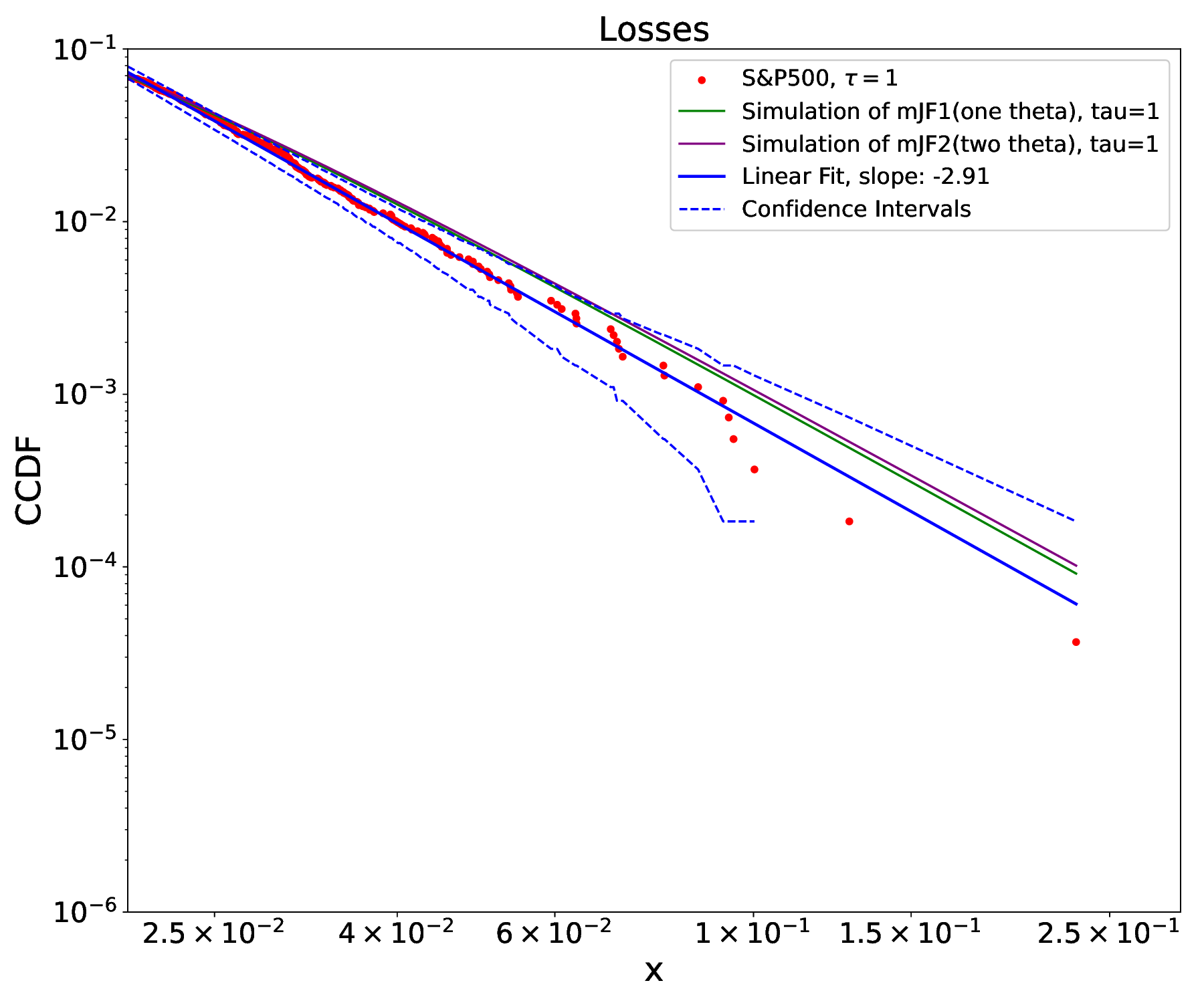}
    \caption{\centering
Linear fit, with its confidence interval, of the tail of the S\&P500 distribution of daily losses; mJF1 and mJF2 fits are shown for comparison.}
    \label{LFCIl}
\end{figure}

\begin{figure}[htbp]
    \centering
    \includegraphics[width=0.7\linewidth]{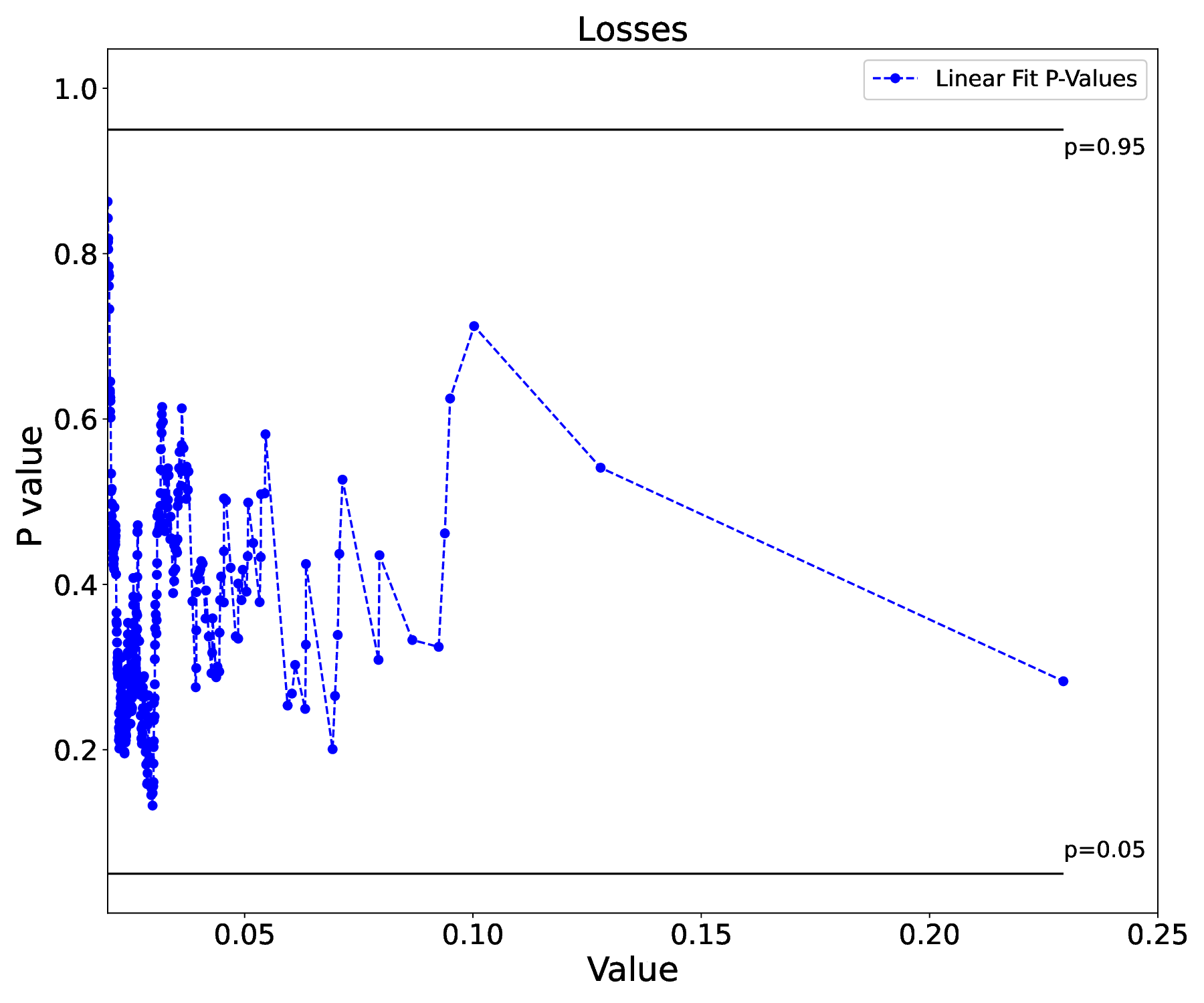}
    \caption{p-values of statistical U-test for the linear fit of the S\&P500 distribution of daily losses.}
    \label{LFPVl}
\end{figure}

\begin{figure}[htbp]
    \centering
    \includegraphics[width=0.7\linewidth]{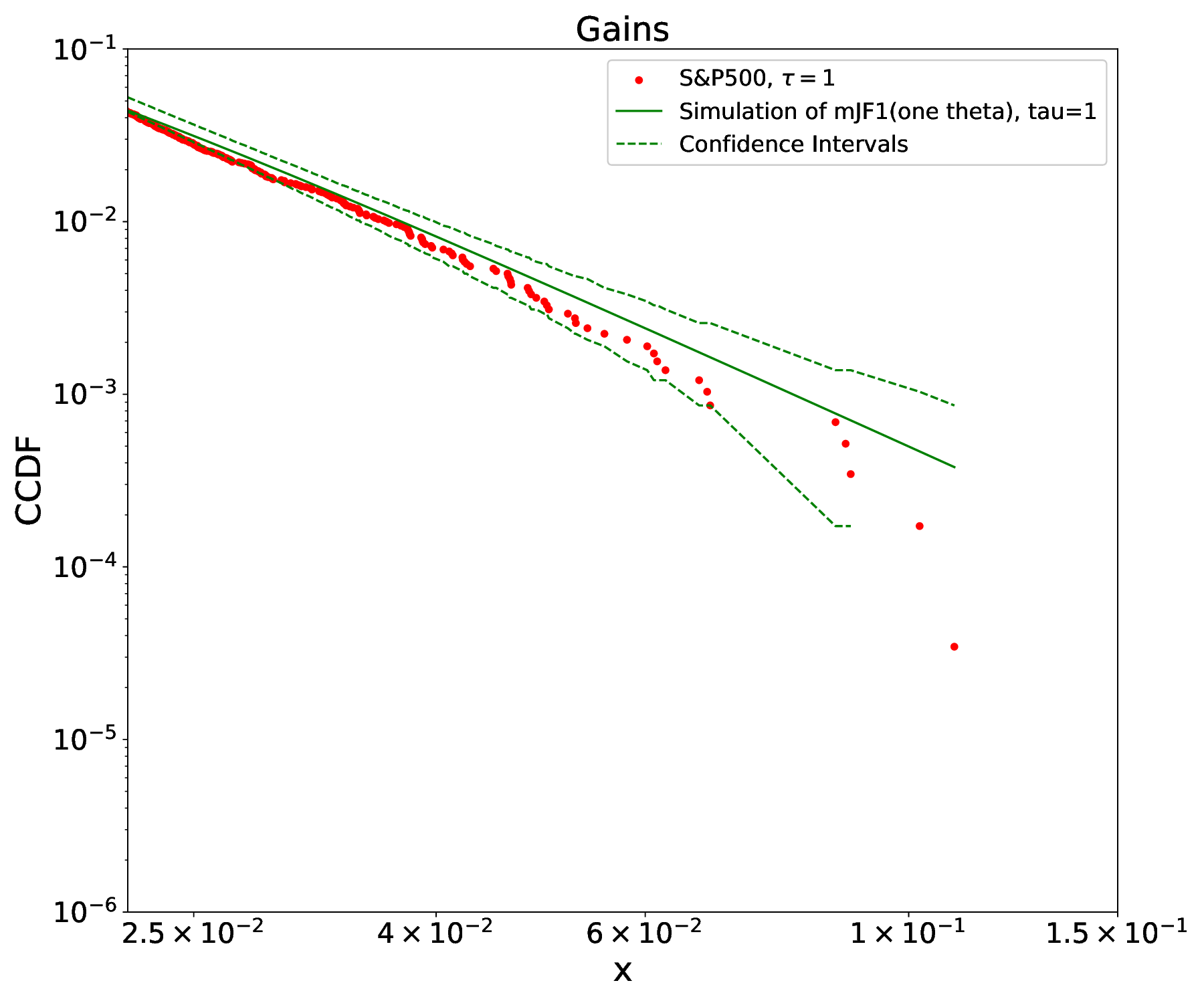}
    \caption{mJF1 fit, with its confidence interval, of the tail of the S\&P500 distribution of daily gains.}
    \label{mJF1CIg}
\end{figure}

\begin{figure}[htbp]
    \centering
    \includegraphics[width=0.7\linewidth]{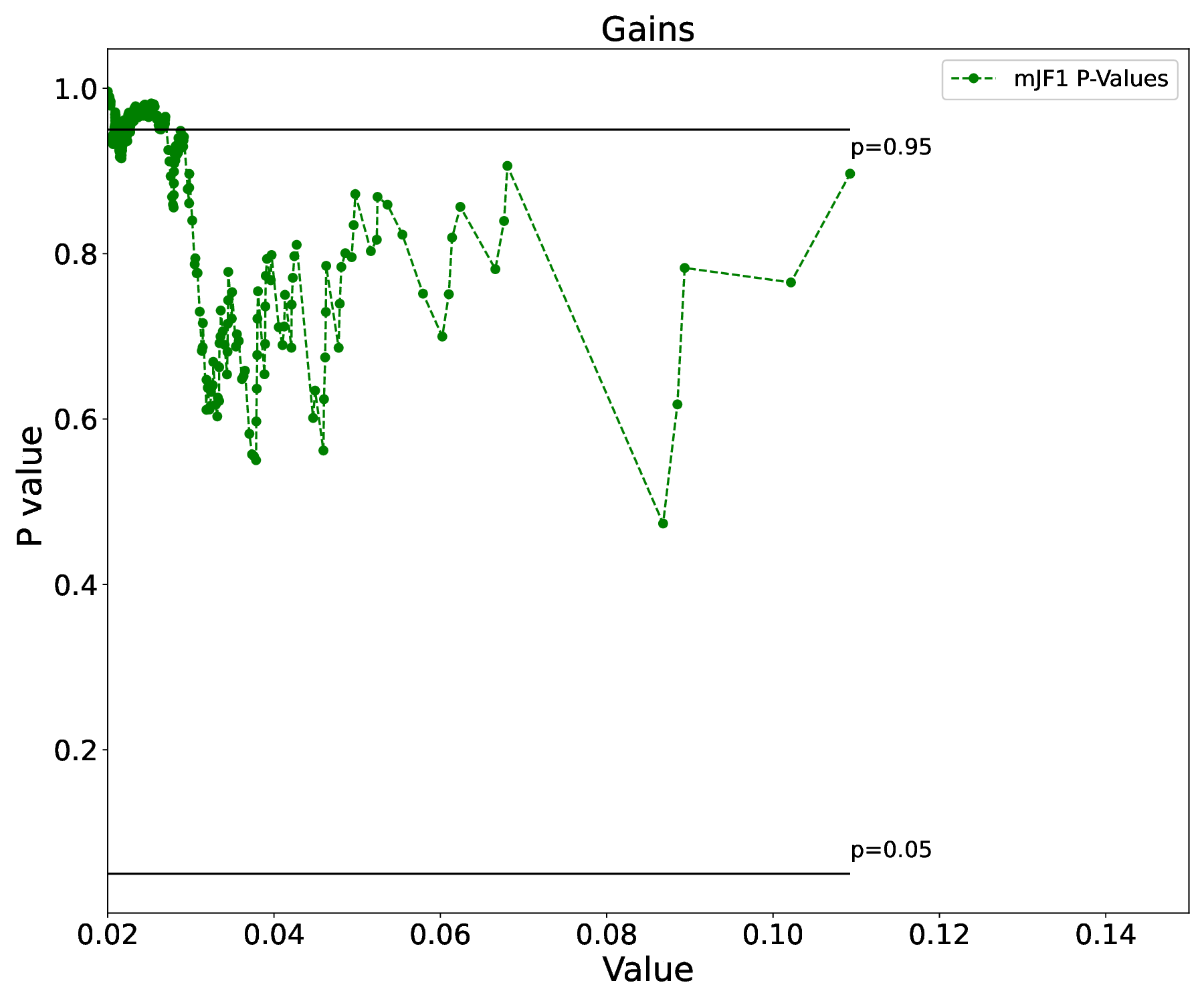}
    \caption{p-values of statistical U-test for mJF1 fit of the S\&P500 distribution of daily gains.}
    \label{mJF1PVg}
\end{figure}

\begin{figure}[htbp]
    \centering
    \includegraphics[width=0.7\linewidth]{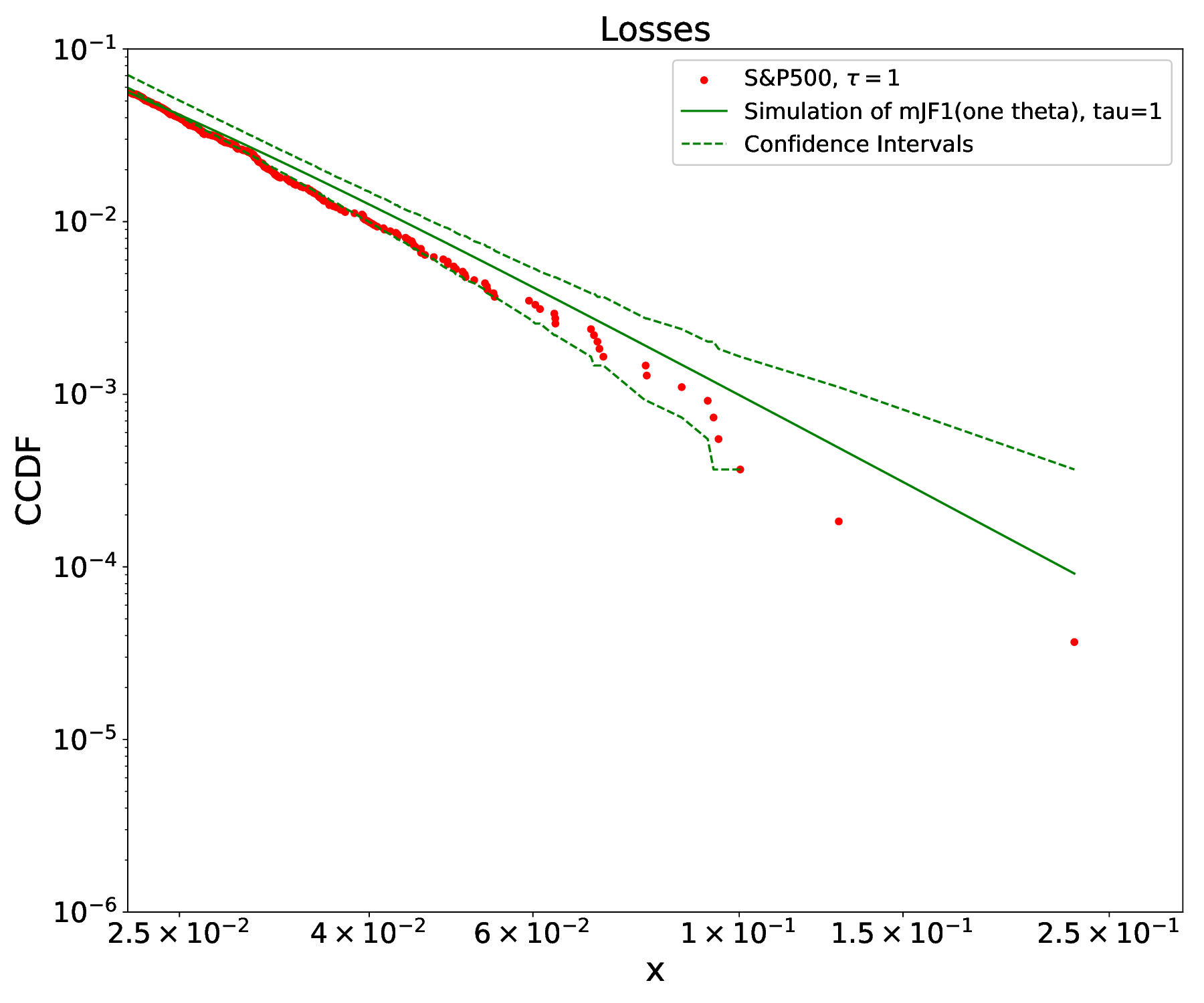}
    \caption{mJF1 fit, with its confidence interval, of the tail of the S\&P500 distribution of daily losses.}
    \label{mJF1CIl}
\end{figure}

\begin{figure}[htbp]
    \centering
    \includegraphics[width=0.7\linewidth]{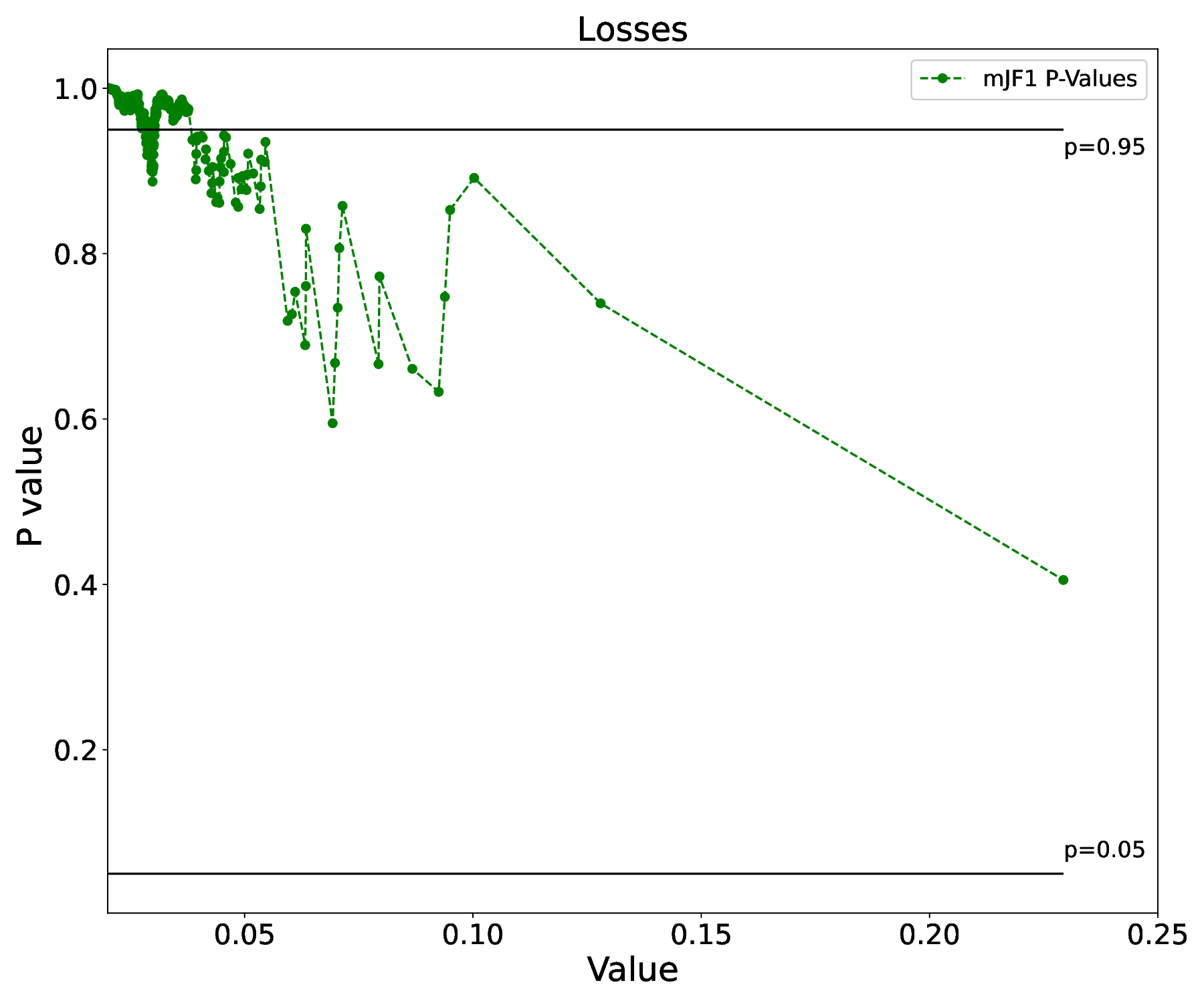}
    \caption{p-values of statistical U-test for mJF1 fit of the S\&P500 distribution of daily losses.}
    \label{mJF1PVl}
\end{figure}

Figs. \ref{LFCIg}-\ref{mJF1PVl} demonstrate the results of statistical tests meant to probe goodness of fit. Figs. \ref{LFCIg} and \ref{LFCIl} show confidence intervals of linear fits and Figs. \ref{mJF1CIg} and \ref{mJF1CIl} show confidence intervals of mJF1 fits. Confidence intervals are obtained using inversion of the binomial distribution \cite{janczura2012black} and we specifically focused on mJF1 as the most transparent and minimal generalization of Student t-distribution. Figs. \ref{LFPVg} and \ref{LFPVl} show p-values of order-statistics-based U-test \cite{pisarenko2012robust} \footnote{We are using eq. (14) in \cite{liu2023dragon} for evaluation of p-values.} for linear fits and Figs. \ref{mJF1PVg} and \ref{mJF1PVl} show p-values for mJF1 fits. It should be noted that both approaches were developed for detections of outliers, such as Dragon Kings and negative Dragon Kings, in the tails of the distributions. For instance values $p<0.05$ and $p>0.95$ would signal a 95\% probability of having, respectively, a Dragon King and a negative Dragon King. In simpler terms, if a data point falls outside the confidence interval and/or if its p-value or (1-p)-value is very small, then it most likely does not belong to the fitting distribution (linear fit being the tail of Pareto distribution). 


\clearpage
\section{Conclusions and Discussion \label{CD}}
The purpose of this work was to glean insight into and to try to analytically describe key empirical findings about S\&P500 1980-2025 returns: heavier tails of losses, leading to the negative skew of the distribution, and positive mean of the distribution, which cannot be entirely attributed to the larger numbers of gains than losses. Our main conclusion is that a modified Jones-Faddy skew t-distribution, (\ref{fmJF1})-(\ref{FmJF1l}) is most likely the best candidate for the stated purpose, even though it is currently unknown how to derive it from first-principles stochastic differential equations.

The main idea behind symmetry breaking of Student t-distribution \ref{fSt}, which is based on (\ref{dxt}) and (\ref{dvt}), is that the latter equation for stochastic volatility is governed by a different set of parameters for gains and losses. In this particular case we operated on the basis of multiplicative model of stochastic volatility (\ref{gvt})-(\ref{alpha}) and so main culprit for explaining heavier power-law dependence of losses is assuming that resulting parameter for losses, $\alpha_l$, is smaller than that for gains, $\alpha_g$, hence the modified Jones-Faddy skew t-distribution mJF1, (\ref{fmJF1}). An additional innocuous introduction of the location parameter helps to explain the positive mean of the distribution. The location parameter seems to account not only for larger number of gains than losses but also for larger values of gains in the bulk of the distribution which offsets the heavier negative tails.

mJF1 still implies that the mean stochastic volatility $\theta$ is the same for gains and losses. To account for the possibility to the contrary we introduced two other distributions with different mean volatilities for gains and losses, $\theta_g$ and $\theta_l$: a mixture half Student-t distribution, (\ref{fhSt})-(\ref{FhStl}) and its simplified form with $\theta_g=\theta_g=\theta$, and the second modified Jones-Faddy distribution mJF2, (\ref{fmJF2})-(\ref{FmJF2l}). The advantage of the former is that it is still rooted in the stochastic differential equation framework. However, due to its structure, it fails to account for the positive mean of actual returns -- daily S\&P500 returns in this case. mJF2, despite and extra parameter, showed virtually no difference relative to mJF1 in fitting the empirical data, both visually and based on statistical tests described in Sec. \ref{NR}. Consequently we believe that mJF1 is the cleanest and most transparent generalization of Student-t for describing daily S\&P500 returns.

There are a number of possible future directions of this work. The most obvious is to consider other market indices. From our previous experience, we expect that DOW will not exhibit significant difference with S\&P500, both in overall behavior and values of parameters. However other long-lasting US indices, such as Russel and NASDAQ, are worth looking into, as well as European and Asian ones. Also overall market returns, reflected by key indices, versus individual companies is a rather challenging question \cite{albuquerque2012skewnwss}. Yet another important avenue is the study of accumulated returns versus dailies: most importantly realized volatility, which shows linear behavior as the function of the number of days of accumulation, and rather abrupt drop off in the tails for longer accumulations. Of course, the most challenging task is finding first-principles explanation of symmetry breaking described in this work.

\section{Data Availability}
We obtained S\&P500 data at Yahoo! Finance. Our datasets are available upon request.
\section{Acknoledgments}
We used MathWorks Matlab for numerical work and Wolfram Mathematica for analytical calculations. Siqi Shao acknowledges support in part by The University of Cincinnati URC Graduate Support Program.
\section{Conflicts of Interest}
The authors declare no conflicts of interest.
 
\newpage
\bibliography{mybib}
\appendix
\section{Derivation of Median \label{median}}
Median is derived from the condition $F(\widetilde{m})=\frac{1}{2}$. For mJF2, for instance, using (\ref{FmJF2g}), this gives 
\begin{equation}
m=\mu+\sqrt{(\alpha_g+\alpha_l)\tau}\;
\frac{2u -1}{\sqrt{1-(2u-1)^2}}
\end{equation}
where 
\begin{equation}
u = I^{-1}\left(\frac{1}{2};\frac{\alpha_l}{\theta_l}+1, \frac{\alpha_g}{\theta_g}+1\right)
\end{equation}
and $I^{-1}$ is the inverse incomplete Beta function. For mJF1 the expression simplifies by setting $\theta_l=\theta_g=\theta$.
\end{document}